\documentclass[twocolumn,aps,showpacs,prb,tightenlines,amsmath,amssymb]{revtex4}
\usepackage{graphicx}
\usepackage{amssymb}
\usepackage{dcolumn}
\usepackage{amsmath}
\usepackage{bm}
\usepackage{colordvi}
\usepackage{mathrsfs}
\makeatletter

\newcommand{\Rmnum}[1]{\expandafter\@slowromancap\romannumeral #1@}
\makeatother

\begin{document}

\title{Fulde-Ferrell-Larkin-Ovchinnikov
  state in spin-orbit-coupled superconductors}  

\author{F. Yang}
\affiliation{Hefei National Laboratory for Physical Sciences at
Microscale, Department of Physics, and CAS Key Laboratory of Strongly-Coupled
Quantum Matter Physics, University of Science and Technology of China, Hefei,
Anhui, 230026, China}

\author{M. W. Wu}
\thanks{Author to whom correspondence should be addressed}
\email{mwwu@ustc.edu.cn.}

\affiliation{Hefei National Laboratory for Physical Sciences at
Microscale, Department of Physics, and CAS Key Laboratory of Strongly-Coupled
Quantum Matter Physics, University of Science and Technology of China, Hefei,
Anhui, 230026, China}

\date{\today}

\begin{abstract} 

We show that in the presence of magnetic field, two
superconducting phases with 
the center-of-mass momentum of Cooper pair parallel to the magnetic field are
induced in spin-orbit-coupled superconductor Li$_2$Pd$_3$B.
Specifically, at small magnetic field, the   
center-of-mass momentum is induced due to the energy-spectrum distortion and no
unpairing region with vanishing singlet correlation appears. We refer to this
superconducting state as the drift-BCS state. By further increasing the magnetic
field, the superconducting state falls into the Fulde-Ferrell-Larkin-Ovchinnikov
state with the emergence of the unpairing regions. The observed abrupt enhancement of the
center-of-mass momenta and suppression on the order parameters during the
crossover indicate the
first-order phase transition. Enhanced Pauli limit and
hence enlarged magnetic-field regime of the Fulde-Ferrell-Larkin-Ovchinnikov
state, due to the spin-flip terms of the spin-orbit coupling, are revealed.
We also address the triplet correlations induced by the spin-orbit coupling, and
show that the Cooper-pair spin polarizations, generated by the magnetic field and
center-of-mass momentum with the triplet correlations, exhibit totally different
magnetic-field dependences between the drift-BCS and
Fulde-Ferrell-Larkin-Ovchinnikov states.

\end{abstract}
\maketitle 

\section{Introduction}

Ever since the Bardeen, Cooper and Schrieffer (BCS) mechanism of
superconductivity was proposed,\cite{BCS}  it is well established that the Cooper pair in 
conventional superconductors such as Al, Pb and Nb, is formed by
two electrons with opposite momenta and spins near the Fermi surface. Together 
with the conventional $s$-wave attractive potential, spatially uniform
singlet order parameter is realized.  After
that, 
possibility of unconventional Cooper pairing together with the 
corresponding nontrivial superconductivity has attracted much
attention. Specifically, in spin space, from the symmetry analysis, triplet
superconductivity with Cooper pair formed by electrons with the same spin
and opposite momenta is theoretically revealed in systems with the broken
space-inversion
symmetry.\cite{TE-0,TE-1,TE-2,TE-3,TE-4,TE-5,TE-6,TE-7,TE-8,TE-10,OP} 
Recently, this possibility has been primarily
realized in noncentrosymmetric superconducting material
Sr$_2$RuO$_4$,\cite{Sr-1,Sr-2,Sr-3,Sr-4,Sr-5,Sr-6,Sr-7,Sr-8} 
and the ongoing experimental evidence makes such material a promising  
candidate to realize non-dissipative spin transport and hence spintronic
application.\cite{SP-1,SP-2,SP-3,SP-4,SP-5,SP-6,SP-7,SP-8,SP-9,SP-10}  

With attention attracted to the orbital degree of freedom of the
pairing, another class of unconventional superconducting 
state, characterized by singlet Cooper pairs with a finite center-of-mass (CM)
momentum, is expected at large magnetic field. This was first predicted by
Fulde and Ferrell (FF)\cite{FF} and a little later by Larkin and Ovchinnikov
(LO)\cite{LO} independently in 1960s. Specifically, the  
presence of the magnetic field leads to the mismatched Fermi surfaces for spin-up
and -down electrons, and consequently, near the corresponding Fermi surfaces,
there exist 
unpairing regions in which the electron can not find the pairing partner with
opposite momentum and spin to form into a Cooper pair. Particularly, when
the magnetic field exceeds a critical strength, by inducing a
finite CM momentum of the Cooper pair, the magnitudes and orientations of the
momenta between the pairing electrons can both be different. Then, the pairing
region between the spin-up and -down electrons is maximized,
leading to the free energy minimized. In this situation, with the
rotational symmetry of the system with respect to the CM-momentum orientation,
FF proposed an order-parameter $\Delta({\bf r})=\Delta_0e^{i{\bf q{\cdot}r}}$ with the
inhomogeneously broadened phase but spatially uniform amplitude\cite{FF} 
whereas LO referred to another order-parameter $\Delta({\bf r})=\Delta_0\cos({\bf
  q{\cdot}r})$ which shows the uniform phase but spatially nonuniform
amplitude.\cite{LO} These two types of the order parameters, now both referred to
as the FFLO state,\cite{FF-od1,FF-od2} have attracted tremendous theoretical and
experimental efforts for decades to prove their existence. Examples include
superconducting heavy fermion\cite{hf-1,hf-2,hf-3,hf-4,hf-5,hf-6,hf-7,hf-8,hf-9}   
and ultracold atom\cite{ca-1,ca-2,ca-3,ca-4,ca-5,ca-6,ca-7,ca-8,ca-9}
systems, multi-band Fe-based superconductors\cite{Fe1,Fe2,Fe3,Fe4} as well as the organic
superconductors.\cite{os-0,os-1,os-2,os-3,os-4,os-5,os-6,os-7}  
However, up till now, the conclusive experimental evidence is still missing.

The experimental difficulty arises from several different aspects. Specifically,
from the FFLO theory, the FFLO state occurs in a very narrow magnetic-field
regime,\cite{FF,FF-od1,FF-od2} leading to the stringent 
experimental requirement.     
The unavoidable disorder in systems  may also destroy the induced CM momentum of Cooper pair and
hence the FFLO state.\cite{hf-8,scatt-1,scatt-2} 
Moreover, in most superconducting materials, the destruction of
superconductivity comes from the orbital effect of the magnetic
field.\cite{FF-od1,FF-od2,Od-1,Od-2} Weak orbital 
depairing effect of the magnetic field is required 
 so that the superconductivity can survive  the transition from
the BCS state into the FFLO one. Consequently, very few 
superconducting materials have been found as possible candidates for the
occurrence of the FFLO state, such as superconducting heavy fermion
material\cite{hf-1,hf-2,hf-3,hf-4,hf-5,hf-6,hf-7,hf-8,hf-9}
and organic superconductors\cite{os-0,os-1,os-2,os-3,os-4,os-5,os-6,os-7}
mentioned above.
Particularly, a magnetic-field-induced superconducting phase in heavy-fermion
compound CeCoIn$_5$ was reported from the thermodynamic
experiments,\cite{hf-1,hf-2,hf-3,hf-4,hf-5,hf-6} which is claimed to be the FFLO state and
more microscopic experimental confirmation is still in progress.

Furthermore, it is reported that the anisotropic Fermi surface is favorable
for the stability of the FFLO state.\cite{ap-1} After that, the
spin-orbit-coupled superconductors attract much attention, since the
interplay between the spin-orbit coupling (SOC)\cite{SOC-1,SOC-2,SOC-3} and
magnetic field can lead to a marked mismatch of the
energy spectra between spin-up and -down electrons as well as the Fermi
surfaces.\cite{ap-2} During the last several years, there indeed
have been several theoretical
reports in spin-orbit-coupled ultracold atomic 
gases\cite{FFs-1,FFs-2,FFs-3,FFs-4,FFs-5,FFs-6,FFs-7,FFs-8} and indirect
experimental evidences from superconducting systems with
SOC\cite{ev1,ev2,ev3,ev4} indicating the
existence for the FFLO state by the  magnetic field. 
Specifically, with SOC and magnetic field, the FFLO
state with induced CM momentum is theoretically predicted by Zheng
{\em et al.}.\cite{FFs-1} Enlarged magnetic-field regime of
the FFLO state due to the enhanced Pauli limit is also revealed in their work,
in accord with the experiments.\cite{ev1,ev2,ev3,ev4} After that, according to
the quasiparticle energy spectra, the FFLO state with SOC is further
divided into gapped and gapless ones, which occur at small and large
magnetic fields, respectively.\cite{FFs-2,FFs-3,FFs-4,FFs-5}
Moreover, it is also reported that
the magnetic field and SOC break the rotational symmetry with respect to the CM-momentum
orientation, and then CM momenta parallel\cite{FFs-2} and
perpendicular\cite{FFs-8} to the magnetic field are predicted in the FFLO states
with Dresselhaus and Rashba SOCs, respectively. Furthermore, by the determined
CM-momentum orientation, it is proposed\cite{FFs-7} that the
FF phase can be enhanced over the LO one. However, the theoretical works above
are based on the numerical calculation of the free-energy minimum with respect to CM momentum and order
parameter, since it is difficult to directly obtain the gap equation by analytically
diagonalizing Hamiltonian with SOC. Therefore, with the numerical difficulty from
multi-variable minimum problem, specific behaviors of the superconducting
state around phase transition are unclear in the literature. The pairing
mechanism and microscopic properties including singlet correlation 
are also beyond this method. Particularly, study of the unpairing regions with
vanishing singlet correlation, which are the hallmark of the FFLO state,\cite{FF} is
still absent.  
    
In this work, we systematically investigate the properties of the FFLO
state with an induced CM momentum in the spin-orbit-coupled superconductors
with the magnetic field. Specifically, by analytically obtaining
the anomalous Green function, we derive the singlet correlation and hence the
gap equation. Then, by self-consistently solving the gap equation, the superconducting state and its
corresponding properties can be determined by numerically calculating the energy
minimum with respect to a single parameter, i.e., the CM momentum.
We further carry out the numerical calculation in superconductor
Li$_2$Pd$_3$B\cite{Li1,Li2,Li3,Li4,Li5,Li6} where strong Dresselhaus SOC\cite{Li2,Li3} and
conventional BCS superconductivity at zero magnetic field\cite{Li1,Li5,Li6} are
realized. 

The calculation shows that with the SOC,
the CM momentum parallel to the magnetic field is induced at small 
field, similar to the gapped FFLO state mentioned
above.\cite{FFs-2,FFs-3,FFs-4,FFs-5} Nevertheless, with an induced CM momentum
in this case, no unpairing region with vanishing singlet correlation is
developed. This is very different from the conventional FFLO state without 
SOC, where the CM momentum is induced simultaneously with the emergence of the
unpairing regions.\cite{FF} By looking into the pairing mechanism, it is further
shown that the 
induced CM momentum with SOC at small magnetic field is due to the energy-spectrum
distortion, resembling the intravalley pairing in
graphene\cite{gi} and transition metal dichalcogenides,\cite{ti} and hence
has different origin from the case in conventional FFLO
state.\cite{FF} Therefore, it is more appropriate to refer to such superconducting
state, in which the CM momentum of the Cooper pair is induced but no
unpairing region is developed, as the drift-BCS state. By further increasing 
the magnetic field, abrupt enhancement of the CM 
momenta and suppression on the order parameters are observed, meaning
the occurrence of the first-order phase transition. Particularly, after the
transition, we find that unpairing regions with vanishing singlet correlation
are induced, indicating the emergence of the FFLO state, resembling the
conventional FFLO one.\cite{FF} We show that the emerged FFLO
state here corresponds to the gapless one mentioned above.\cite{FFs-2,FFs-3,FFs-4,FFs-5}
Enhanced Pauli limit and hence enlarged magnetic-field regime of the emerged
FFLO state by SOC are also observed in our work. We further show that the
enhancement of the Pauli limit is due to the spin-flip terms of the SOC, which
suppress the unpairing regions. Finally, we discuss the triplet correlations induced by the 
SOC,\cite{TE-2,TE-5,TE-10,OP} and show that the Cooper-pair spin
polarizations,\cite{TE-0,TE-1,csp1,csp2} which are predicted to be induced by
magnetic field and CM momentum in the presence of triplet
correlations,\cite{csp2} exhibit totally different magnetic-field dependences between the
drift-BCS and FFLO states. This provides 
an experimental scheme to distinguish these two phases through the reported 
magnetoelectric Andreev effect,\cite{csp1,csp2,csp3} in addition to the phase
transition. 

This paper is organized as follows. In Sec. II, we introduce
our model and present the calculation of the energy for superconducting
state. The specific numerical results in Li$_2$Pd$_3$B and analytic analysis are
presented in Sec. III. We summarize in Sec. IV.

\section{MODEL}
\label{model}

In this section, we first present the Hamiltonian of the spin-orbit coupled $s$-wave
superconductor in the presence of the magnetic field. Then we give the gap
equation and lay out the calculation of the energy for the superconducting
state. 

\subsection{HAMILTONIAN AND  GAP EQUATION}

With the magnetic field and CM momentum of Cooper pair, by
defining the Nambu spinors {\small {${\hat \Phi}_{\bf
  k}=[\phi_{\uparrow{\bf k+q}},\phi_{\downarrow{\bf 
    k+q}},\phi^{\dagger}_{\uparrow{\bf -k+q}},\phi^{\dagger}_{\downarrow{\bf
  -k+q}}]^{T}$}}, we present the
Hamiltonian $H_{S}$ of the spin-orbit-coupled $s$-wave superconductor as:\cite{FF,OP,csp2} 
\begin{equation}
\label{Hamiltonian}
{\hat H_{S}}=\frac{1}{2}\int{d{\bf k}}{\hat \Phi}^{\dagger}_{\bf
  k}{\hat H_s}({\bf k})\rho_3{\hat \Phi}_{\bf k},
\end{equation}
with
\begin{eqnarray}
&&{\hat H_{s}}({\bf k})=\left(\begin{array}{cc}
\xi_{\bf k^+}+{\bf \Omega}_{\bf k^+}\cdot{\bm \sigma} & \Delta_{\bf q}i\sigma_2\\
\Delta^*_{\bf q}i\sigma_2 & \xi_{\bf k^-}+{\bf \Omega}_{\bf k^-}\cdot{\bm
\sigma}\\
\end{array}\right).~~~~~
\end{eqnarray}
Here, $\rho_3=\sigma_0\otimes\tau_3$; $\sigma_i$ and $\tau_i$ stand for the Pauli
matrices in spin and particle-hole spaces, respectively; ${\bf k}^{\pm}=\pm{\bf
  k+q}$ with ${\bf q}$ standing for the CM momentum; $\xi_{\bf k}=\varepsilon_{\bf k}-E_F$ and 
$\varepsilon_{\bf k}=k^2/(2m^*)$ with $m^*$ being the 
effective mass of electrons in superconductor and $E_F$ denoting the Fermi
energy; ${\bf \Omega}_{\bf k}={\bf  
  h_k}+{\bf h}_B$ with ${\bf h}_{\bf k}$ and ${\bf h}_B$ representing the SOC 
and Zeeman energy, respectively; $\Delta_{\bf q}=-V\sum'_{\bf k}\langle{\phi_{\uparrow{\bf
      k+q}}\phi_{\downarrow{\bf -k+q}}}\rangle$ stands for the order parameter
of the superconducting state in the momentum space; 
$\langle~\rangle$ stands for the ensemble average; $V$ is the conventional $s$-wave
attractive potential in superconductors. It is noted that the order parameter at
this case can be transformed into the FF form\cite{FF} $\Delta({\bf
  r})=\Delta_{\bf q}e^{i{\bf q{\cdot}r}}$ in real space. 

In Nambu$\otimes$spin space, the equilibrium Green function\cite{G1,G2,G3}
in the momentum 
space is given by 
\begin{equation}
G_{\bf q}({\bf k},{\tau})=-\rho_3\langle{T_{\tau}{\hat \Phi}_{\bf
  k}(\tau){\hat \Phi}^{\dagger}_{\bf
  k}(0)}\rangle, 
\end{equation}
where $T_{\tau}$ represents the chronological-ordering operator; $\tau$ is
the imaginary time. By expressing 
\begin{equation}
G_{\bf q}({\bf k},\tau)=\left(\begin{array}{cc}
g_{\bf q}({\bf k},\tau)& f_{\bf q}({\bf k},\tau)\\ 
f^{\dagger}_{\bf q}({\bf -k},\tau)& {g^{\dagger}}_{\bf q}({\bf -k},\tau)\\ 
\end{array}\right),
\end{equation} 
one can obtain the normal Green function $g_{\bf q}({\bf k},\tau)$ and anomalous Green
function $f_{\bf q}({\bf k},\tau)$.\cite{TE-2,G1,G2,G3}

Then, in the Matsubara
representation\cite{G1,G2,G3} $G_{\bf
  k}(i\omega_n)=\int^{\beta}_{0}d{\tau}e^{i\omega_n{\tau}}G_{\bf k}(\tau)$, from the Gor'kov
equation:\cite{Gorkov}
\begin{equation}
\label{frequency}
[i\omega_n\rho_3-H_{s}({\bf k})]G_{\bf q}({\bf k},i\omega_n)=1,
\end{equation}
one has
\begin{equation}
  \label{pro1}
(i\omega_n-\xi_{\bf k^+}-{\bf \Omega}_{\bf k^+}\cdot{\bm \sigma})f_{\bf q}({\bf
    k},i\omega_n)-\Delta_{\bf q}i\sigma_2g^{\dagger}_{\bf q}({\bf
    k},i\omega_n)=0,
\end{equation}
\begin{equation}
  \label{pro2}
-\Delta^*_{\bf q}i\sigma_2f_{\bf q}({\bf
    k},i\omega_n)+(i\omega_n-\xi_{\bf k^-}-{\bf \Omega}_{\bf k^-}\cdot{\bm
\sigma})g^{\dagger}_{\bf q}({\bf
    k},i\omega_n)=1.  
\end{equation}
Here, $\beta=1/(k_BT)$ with $k_B$
being the Boltzmann constant and $T$ representing the temperature;
$\omega_n=(2n+1)\pi{k_BT}$ are the Matsubara frequencies with $n$ being integer.

Following the previous work,\cite{csp2} through multiplying
Eq.~(\ref{pro2}) by $i\sigma_2$ from the left side, one
immediately has
\begin{equation}
  \label{pro3}
\Delta_{\bf q}^*f_{\bf q}({\bf
    k},i\omega_n)+(i\omega_n-\xi_{\bf k^-}+{\bf \Omega}_{\bf k^-}\cdot{\bm
\sigma}^*)i\sigma_2g^{\dagger}_{\bf q}({\bf
    k},i\omega_n)=i\sigma_2.
\end{equation}
Then, by using Eq.~(\ref{pro1}) to replace $i\sigma_2g^{\dagger}_{\bf q}({\bf k},i\omega_n)$ in Eq.~(\ref{pro3}),  the
anomalous Green function can be obtained:  
\begin{equation}
f_{\bf q}({\bf k},i\omega_n)=[f^s_{\bf q}({\bf k},i\omega_n)+{\bf f}^t_{\bf
  q}({\bf k},i\omega_n)\cdot{\bm \sigma}]i\sigma_2,
\end{equation}
with the singlet $f^s_{\bf q}({\bf k},i\omega_n)$ and triplet ${\bf f}^t_{\bf
  q}({\bf
  k},i\omega_n)=(\frac{f_{\downarrow\downarrow}-f_{\uparrow\uparrow}}{2},\frac{f_{\downarrow\downarrow}+f_{\uparrow\uparrow}}{2i},f_{\downarrow\uparrow+\uparrow\downarrow})$
pairings\cite{tp1,tp2} 
written as:  
\begin{eqnarray}
f^s_{\bf q}({\bf k},i\omega_n)&=&\frac{\Delta_{\bf q}}{w_{\bf q}({\bf k},i\omega_n)}[(i\omega_n-\xi_{\bf k^+})(i\omega_n+\xi_{\bf
  k^-})\nonumber\\
&&\mbox{}+{\bm \Omega}_{\bf k^+}\cdot{\bm \Omega}_{\bf k^-}-|\Delta_{\bf q}|^2],\\
{\bf f}^t_{\bf q}({\bf k},i\omega_n)&=&\frac{\Delta_{\bf q}}{w_{\bf q}({\bf
    k},i\omega_n)}[(i\omega_n-\xi_{\bf k^+}){\bm \Omega}_{\bf 
  k^-}+(i\omega_n\nonumber\\
&&\mbox{}+\xi_{\bf 
  k^-}){\bm \Omega}_{\bf k^+}-i{\bm \Omega}_{\bf k^+}\times{\bm \Omega}_{\bf k^-}],\label{tp}\\
w_{\bf q}({\bf k},i\omega_n)&=&{\prod_{\mu=\pm}(i\omega_n-E^e_{\mu{\bf k}})(i\omega_n-E^h_{\mu{\bf k}})}.
\end{eqnarray}
$E^{e(h)}_{\mu{\bf k}}$ ($\mu=\pm$) stand for the quasiparticle electron (hole)
energy spectra in superconductors, which can be obtained from the solutions
of equation $|f^s_{\bf q}({\bf
  k},\omega)|^2-|{\bf f^{t}_q}({\bf
  k},i\omega)|^2=0$ with respect to $\omega$.

With the anomalous Green function, the singlet $\rho^s_{\bf q}$ and triplet ${\bm
  \rho}_{\bf
  q}^t=(\frac{{\rho}^t_{s=-1}-{\rho}^t_{s=1}}{2},\frac{{\rho}^t_{s=-1}+{\rho}^t_{s=1}}{2i},{\rho}^t_{s=0})$\cite{tp1,tp2,tp3}  
correlations are defined as
\begin{eqnarray}
\label{scorrelation}
\rho^s_{\bf q}({\bf k})&=&-\frac{1}{\beta}\sum_{i\omega_n}f^s_{\bf q}({\bf
  k},i\omega_n),\\
{\bm \rho}_{\bf q}^t({\bf k})&=&-\frac{1}{\beta}\sum_{i\omega_n}{\bf f}^t_{\bf q}({\bf
  k},i\omega_n).\label{tcorrelation}
\end{eqnarray}
Then one immediately has the gap equation:\cite{Gorkov} 
\begin{equation}
\label{GE}
{\Delta_{\bf q}}=V{\sum_{{\bf k}}}'\rho^s_{\bf q}({\bf k}).
\end{equation} 
Here, the summation is taken for the values of
${\bf k}$ satisfying $|E_{\uparrow{\bf k}}-E_F|<\omega_D$ and $|E_{\downarrow{\bf
    k}}-E_F|<\omega_D$ where $E_{\uparrow(\downarrow){\bf k}}$ is the energy of
spin-up (-down) electron with the momentum ${\bf k}$; $\omega_D$ stands for
the Debye frequency. It is noted that due to the conventional $s$-wave
attractive potential, only singlet order parameter exists.
Then, by self-consistently solving
Eq.~(\ref{GE}), the order parameter ${\Delta_{\bf q}}$ at fixed CM momentum of Cooper
pair ${\bf q}$ is obtained.

\subsection{GROUND-STATE ENERGY}

Following the previous work by FF,\cite{FF} by replacing the $s$-wave  
attractive potential $V$ with an effective one $\lambda$ in
Eq.~(\ref{GE}), an potential-dependent order
parameter $\Delta_{\bf q}(\lambda)$ can be immediately obtained by self-consistently solving 
Eq.~(\ref{GE}).  

Then, by neglecting the Fock energy of the normal state, based on Feynman-Hellmann theorem
\begin{equation}
\partial_{\lambda}E=\langle\partial_{\lambda}H_s(\lambda)\rangle=-\frac{|\Delta_{\bf
  q}(\lambda)|^2}{\lambda^2},
\end{equation}
the expectation value of the
energy difference between the superconducting state $E^S_{\bf q}$ and
normal one $E^N$ is given by 
\begin{equation}
\label{FE}
\delta{E}_{\bf q}=E^S_{\bf q}-E^N=-\int^{V}_{V_0}d\lambda\frac{|\Delta_{\bf
  q}(\lambda)|^2}{\lambda^2},
\end{equation} 
where $V_0$ is the effective attractive potential at the transition point
between the superconducting state and the normal one [$\Delta_{\bf q}(V_0)=0$].
Then, by calculating the minimum of ${\delta}E_{\bf q}$ with respect
to a single parameter, i.e., the CM momentum ${\bf q}$, the properties of the
superconducting state including the CM momentum, order parameter, quasiparticle
energy spectra, singlet and triplet correlations can all be determined.
Particularly, it is noted that through self-consistently solving the gap
equation in our work, the numerical difficulty from the multi-variable minimum
problem mentioned in the introduction is reduced, leading to more accurate
results.

\begin{table}[htb]
  \caption{Parameters used in our calculation. Note that $m_0$ stands for the
    free electron mass. $\Delta_0$ is the order parameter at $T=0~$K without
    magnetic field. $V$ is obtained by fitting $\Delta_0$. $k_F$ is the largest
    momentum with the Fermi energy in the absence of the magnetic field.}  
  \label{parametertable} 
  \begin{tabular}{l l l l}
    \hline
    \hline
    $m^*/m_0$&\quad$3.049^a$&\quad\quad$E_F~$(meV)&\quad$70^a$\\ 
    $\Delta_0$~(meV)&\quad$1.2732^b$&\quad\quad$\omega_D~$(meV)&\quad$38.088^b$\\
    $\gamma~$(meV${\cdot}${\AA})&\quad$23.28^a$&\quad\quad$T~$(K)&\quad$0$\\
    $V$~(meV)&\quad$0.218$&\quad\quad$k_F~$({\AA}$^{-1}$)&\quad$0.246$\\
    \hline
    \hline
  \end{tabular}\\
   \quad$^a$ Refs.~\onlinecite{Li1}. \quad$^b$ Ref.~\onlinecite{Li2}.
\end{table}

\section{NUMERICAL RESULTS}

In this section, by numerically solving Eq.~(\ref{FE}), we discuss the
properties of the superconducting state in spin-orbit-coupled superconductors. 
The specific numerical calculation is carried out in the material
Li$_2$Pd$_3$B,\cite{Li1,Li2,Li3,Li4,Li5,Li6} in which the conventional s-wave BCS
behavior\cite{Li2,Li3} and strong SOC\cite{Li1,Li5,Li6} have been realized. The
SOC in Li$_2$Pd$_3$B has the Dresselhaus form ${\bf h}_{\bf k}=\gamma{\bf
  k}$,\cite{Li5,Li6} in consistence with the cubic symmetry. All the material
parameters used in our calculation are listed in Table~\ref{parametertable}. The
magnetic field is chosen along the $z$ direction.

\begin{figure}[htb]
  {\includegraphics[width=8.5cm]{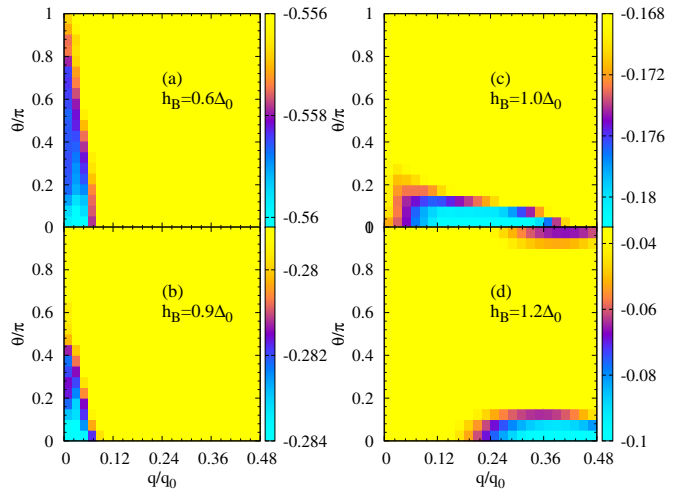}}
\caption{$\delta{E}_{\bf q}$ versus longitude $\theta$ (with respect to ${\bf
    h}_B$) and magnitude $q$ of the CM momentum at (a) $h_B=0.6\Delta_0$, 
(b) $h_B=0.9\Delta_0$, (c) $h_B=1.0\Delta_0$ and (d)
$h_B=1.2\Delta_0$. $q_0=\sqrt{2m^*(E_F+h_B)}-\sqrt{2m^*(E_F-h_B)}$.}    
\label{figyw1}
\end{figure}

\subsection{Orientation of CM momentum and Pairing Mechanism}
\label{od}

We first focus on the CM-momentum dependences of the energy of the
superconducting state at different magnetic fields. We find that the energy difference
$\delta{E}_{\bf q}$ between the superconducting state and normal one always
shows isotropy with respect to the latitude of the CM momentum around the magnetic field (not
shown). This is due to the spatial-rotational symmetry around the magnetic field
of the system. Then, $\delta{E}_{\bf q}$ as function of the longitude $\theta$
(with respect to the
magnetic field) and magnitude $q$ of the CM momentum are plotted in
Fig.~\ref{figyw1} at different magnetic fields. As seen from the figure,
in the presence of the magnetic field, the minimum of $\delta{E}_{\bf q}$ is always reached at
finite $q$ with $\theta=0$ ($z$ direction). This indicates that a CM momentum 
 parallel to magnetic field is induced in the
superconducting state in the presence of the magnetic field and SOC, similar to
the previous work with the same Dresselhaus SOC.\cite{FFs-2} Small and large CM momenta are observed before
[Figs.~\ref{figyw1}(a) and (b)] and after 
[Figs.~\ref{figyw1}(c) and (d)] $h_B=\Delta_0$ here, respectively. 
Particularly, with the
determined orientation according to the energy minimum, the induced CM
momentum is inherently robust against the impurity scattering.  

\begin{figure}[htb]
  {\includegraphics[width=8.5cm]{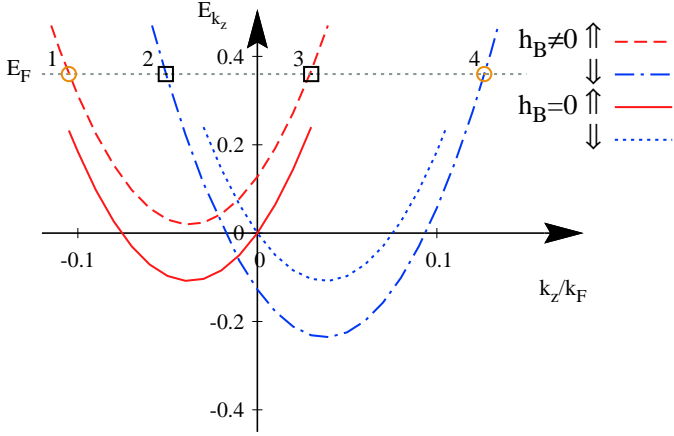}}
\caption{Energy spectra of spin-up and -down electrons along $k_z$
  direction. Dashed and chain curves: SOC and magnetic field are both included;
  Solid and dotted curves: only SOC in included. The horizontal dotted line indicates
  the Fermi surface. Circles (Squares): type I (II) pairing formed by spin-up
  electron 1 (3) and spin-down one 4 (2).}    
\label{figyw2}
\end{figure}

To illustrate the pairing mechanism, we further plot the energy spectra of
spin-up and -down electrons along the $k_z$ direction (magnetic-field direction)
with and without magnetic field in Fig.~\ref{figyw2}. As
seen from the figure, without the magnetic field, the SOC leads to the opposite
shifts of the energy spectra between spin-up (red solid curve) and -down (blue dotted curve)
electrons along the $k_z$ direction. Then, the magnetic field causes the
opposite energy shifts of the energy spectra between spin-up (red dashed curve) and
-down (blue chain curve) electrons. In this situation, by assuming the Debye
frequency $\omega_D=0$, the Cooper pairing only occurs at the Fermi surface, as
shown in Fig.~\ref{figyw2}. Then, 
there exist two possible types of Cooper pairings: type I,
formed by spin-up electron 1 (with $-{\bf k}_{\rm I}+{\bf q}$) 
and spin-down one 4 (with ${\bf k}_{\rm I}+{\bf q}$) in 
Fig.~\ref{figyw2} in favor of the CM momentum ${\bf q}=q_c{\bf z}$
($q_c=\frac{\sqrt{|m^*\gamma|^2+2m^*(E_F+h_B)}-\sqrt{|m^*\gamma|^2+2m^*(E_F-h_B)}}{2}$); 
type II, formed by spin-up electron 3 (with ${\bf k}_{\rm 
    II}+{\bf q}$) and spin-down one 2 (with $-{\bf k}_{\rm
    II}+{\bf q}$) in Fig.~\ref{figyw2}, in favor of the CM momentum ${\bf
    q}=-q_c{\bf z}$. Nevertheless, from Fig.~\ref{figyw1}, the CM momentum ${\bf
  q}$ is along the ${\bf z}$ direction, as mentioned above. This indicates the type
I pairing makes the leading contribution in the determination of the CM momentum. 

The leading role of type I pairing can be understood as follows. On one hand, 
the relative momentum $k_{\rm I}$ in type I pairing is
larger than $k_{\rm II}$ in type II one. Then, with the larger relative momentum
$k$ and hence larger density of states in Eq.~(\ref{GE}), type I pairing makes
the leading contribution to the summation with respect to the momentum space. 
On other hand, from the framework 
of the Ginzburg-Landau theory, the free energy densities of superconducting
system reads
\begin{equation}
\label{freeenergy}
\mathcal{F}=\alpha|\psi({\bf r})|^2+\frac{\eta}{2}|\psi({\bf r})|^4+\frac{1}{2m}[{\bf
    \Pi}\psi({\bf r})]^*[{\bf \Pi}\psi({\bf r})],
\end{equation}
with ${\bf \Pi}=-i\hbar{\bm \nabla}+2e{\bf A}$ and $\psi({\bf r})=\Delta({\bf
  r})/V$; $\alpha$ and $\eta$ being the 
corresponding expansion parameters; ${\bf A}$ standing for the magnetic vector
potential. In the presence of the SOC, one can replace ${\bf A}$ by ${\bf A_s}$
with ${\bf A_s}=\gamma{\bf s}$ for the Dresselhaus SOC and ${\bf s}$ representing
the spin vector of electrons. Then, terms related to the SOC in
Eq.~(\ref{freeenergy}) are written as
\begin{eqnarray}
\label{FS}
\mathcal{F}_{s}&=&\frac{{\rm Re}[\Delta^*({\bf r})2e{\bf A}_s\cdot(-i\hbar{\bm
    \nabla})\Delta({\bf r})]}{mV^2}+\frac{e^2|{\bf A}_s\Delta({\bf r})|^2}{mV^2}\nonumber\\
&=&\frac{2e\hbar\gamma|\Delta_{\bf
    q}|^2}{mV^2}{\bf s{\cdot}q}+\frac{2e^2\gamma^2|{\bf s}\Delta_{\bf q}|^2}{mV^2},
\end{eqnarray} 
where we have applied $\Delta({\bf r})=\Delta_{\bf q}e^{i{\bf q{\cdot}r}}$ in
Eq.~(\ref{FS}). With the magnetic field, the spin vector ${\bf s}$
is anti-parallel to ${\bf h}_B$. Hence, to obtain the free-energy
minimum, the induced CM momentum ${\bf q}$ should be parallel to the magnetic
field (${\bf z}$ direction), in accord with the type I pairing (in favor of ${\bf q}=q_c{\bf z}$).   

\begin{figure}[htb]
  {\includegraphics[width=9.0cm]{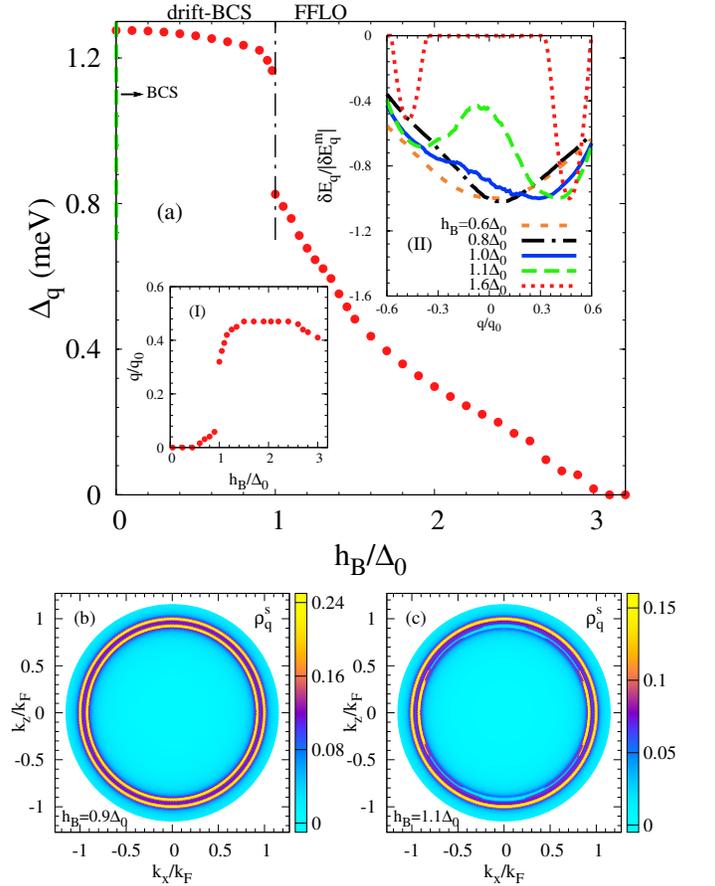}}
\caption{(a): magnetic-field dependence of ${\Delta_{\bf q}}$. The vertical
  dashed (chain) line indicates the BCS state (crossover between the drift-BCS and
  FFLO states). The inset (I) in (a) shows the CM 
  momentum as function of magnetic field. The inset (II) in (a) exhibits $\delta{E}_{\bf
    q}$ versus ${\bf q}=q{\bf z}$ at different magnetic fields. The results in
  inset (b) is renormalized by the maximum of $|\delta{E}_{\bf
    q}|$ for each magnetic field. $q_0=\sqrt{2m^*(E_F+h_B)}-\sqrt{2m^*(E_F-h_B)}$. (b) [(c)]: singlet
  correlations in the momentum space at $h_B=0.9\Delta_0$ ($h_B=1.1\Delta_0$). }   
\label{figyw3}
\end{figure}

\subsection{Phase diagram}
\label{pd}
In this part, we discuss the phase diagram of the superconducting state. The magnetic
field dependences of the order parameter $\Delta_{\bf q}$ and
CM momentum ${\bf q}=q{\bf z}$ are plotted in Fig.~\ref{figyw3}(a) and the inset
(I) of Fig.~\ref{figyw3}(a), respectively. In the calculation, ${\bf q}$ is
chosen at the minimum of $\delta{E_{\bf q}}$. As seen from Fig.~\ref{figyw3}(a)
and inset (I), with the increase of magnetic field from zero, before
reaching $h_B=\Delta_0$, the order parameter decreases slightly
[Fig.~\ref{figyw3}(a)]. In the same time, a CM momentum is induced and increases
from zero [inset (I)], similar to the gapped FFLO state in the previous
    works\cite{FFs-2,FFs-3,FFs-4,FFs-5} mentioned in the
    introduction. Nevertheless, by plotting the singlet correlation at 
$h_B=0.9\Delta_0$ in Fig.~\ref{figyw3}(b), it is seen from the figure that two 
separated and complete circles with finite $\rho^s_{\bf q}({\bf k})$,
corresponding to type I (large $k$) and II (small $k$) pairings, are observed
due to the presence of the SOC, and no unpairing region with vanishing
$\rho^s_{\bf q}({\bf k})$ appears when $h_B<\Delta_0$. As mentioned in the
introduction, it is more appropriate to refer to such superconducting
state in which the CM momentum is induced but no unpairing region is observed,
as the drift-BCS state, since the induced CM momentum at small magnetic field
here arises from the energy-spectrum 
distortion by the magnetic field and SOC as mentioned in
Sec.~\ref{od}, resembling the intravalley pairing in 
graphene\cite{gi} and transition metal dichalcogenides,\cite{ti} and hence
has different origin from the case in the conventional FFLO
state without SOC.\cite{FF}. Particularly,
since this drift-BCS state occurs at small magnetic field, it can inherently
survive against the orbital depairing effect. 
 
By further increasing the magnetic field, abrupt suppression on the
order parameters [shown in Fig.~\ref{figyw3}(a)] and enhancement of the CM
momenta [shown in 
the inset (I) of Fig.~\ref{figyw3}(a)] are observed before and after
$h_B\approx\Delta_0$, indicating the first-order phase transition at the
crossover. The abrupt changes can be understood from the inset (II) of 
Fig.~\ref{figyw3}(a) where we plot the energy differences
$\delta{E}_{\bf q}$ versus
${\bf q}=q{\bf z}$ at different magnetic fields. From the inset (II), it is seen
that when $h_B<\Delta_0$, the minimum of $\delta{E}_{\bf q}$ sits near
$q\approx0$ (brown dashed curve), and the minimum position $q_m$
increases with the magnetic field by comparing the brown dashed curve at
$h_B=0.6\Delta_0$ with the black chain one at $h_B=0.8\Delta_0$. 
Whereas with the increase of magnetic
field when $h_B\geq\Delta_0$, $\delta{E}_{\bf q\approx0}$ increases and
then the minimum of $\delta{E}_{\bf q}$ appears around $q\approx0.45q_0$,
leading to abrupt changes of the order parameters and CM momenta before and
after $h_B\approx\Delta_0$. Moreover, we plot the singlet correlation when
$h_B>\Delta_0$ in Fig.~\ref{figyw3}(c). In comparison with the two complete
circles of singlet correlations at $h_B<\Delta_0$ [Fig.~\ref{figyw3}(b)], in the 
case with $h_B>\Delta_0$ [Fig.~\ref{figyw3}(c)], the inner circle (type II
pairing) is broken around $k_z$ axis whereas the
outer circle (type I pairing) survives since the CM momentum is along the
favorable orientation to type I pairing (Sec.~\ref{od}). The appeared 
regions with the destroyed singlet correlations in this situation, known as the
unpairing ones, are the hallmark of the emergence of the FFLO state.\cite{FF} 

\begin{figure}[htb]
  {\includegraphics[width=9.0cm]{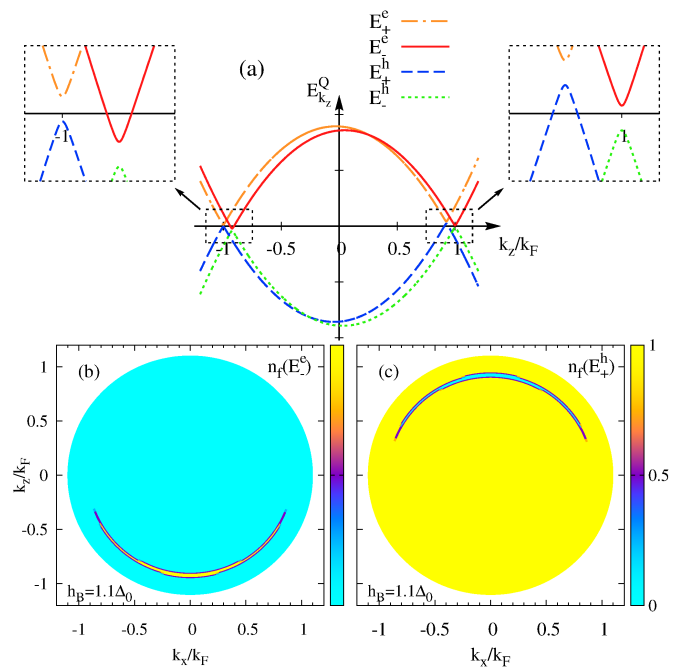}}
  \caption{(a): quasiparticle electron and hole energy spectra along $k_z$ direction;
    (b) [(c)]: distribution of the quasiparticle electron $n_f(E^e_{-{\bf
        k}})$ [hole $n_f(E^h_{+{\bf k}})$] in the momentum space at $h_B=1.1\Delta_0$. It is also
    found that $n_f(E^e_{+{\bf
        k}})=0$ and $n_f(E^h_{-{\bf k}})=1$ in the momentum space.}   
\label{figyw4}
\end{figure}

The destruction mechanism of the singlet correlation for the inner circle around $k_z$
axis can be understood as follows in a special direction. With the
induced CM momentum pointing to $k_z$ direction (in
Sec.~\ref{od}), along $k_z$, one has ${\bf h}_{\bf
      k}=\gamma(k_x,k_y,k_z)=\gamma(0,0,k_z)$ where the spin-flip terms of the
    SOC are zero, and then
the singlet correlation can be simplified into
\begin{equation}
\label{sc}
\rho^s_{\bf q}({\bf k})={\rm Tr}\left\{\frac{2\Delta_{\bf q}[n_f(E^h_{\sigma_3{\bf k}})-n_f(E^e_{\sigma_3{\bf k}})]}{\sqrt{(\xi_{k}+\sigma_3\gamma{k_z})^2+|\Delta_{\bf
    q}|^2}}\right\}. 
\end{equation}
Here, the quasiparticle electron and hole energy spectra are written as: 
\begin{eqnarray}
&&E^{e}_{\sigma_3{k{\bf z}}}=\sqrt{(\xi_{k}+\sigma_3\gamma{k_z})^2+|\Delta_{\bf
    q}|^2}+\frac{k_zq}{m^*}+\mu{\Omega_{q{\bf z}}},\label{ee}\\ 
&&E^{h}_{\sigma_3{k{\bf z}}}=-\sqrt{(\xi_{k}+\sigma_3\gamma{k_z})^2+|\Delta_{\bf
    q}|^2}+\frac{k_zq}{m^*}+\mu{\Omega_{q{\bf z}}}.~~~~~\label{he}
\end{eqnarray}
which are plotted Fig.~\ref{figyw4}(a). When $k_z>0$, the
maximum of factor ${\Delta_{\bf q}}/{\sqrt{(\xi_{k}+\sigma_3\gamma{k_z})^2+|\Delta_{\bf
    q}|^2}}$ occurs at large (small) $k_z$ for $\sigma_3=-1$ ($\sigma_3=1$) in
the summation of Eq.~(\ref{sc}). Hence,
$\sigma_3=-1$ ($\sigma_3=1$) makes the main contribution to the outer
(inner) circle of the singlet correlation, e.g., type I (II) pairing, in
accord with the pairing spin-down electron 4 (spin-up one 3) in
Fig.~\ref{figyw2}. In this case, for $\sigma_3=-1$, as shown in
Fig.~\ref{figyw4}(a), one always has $E^e_->0$ (red solid curve) and $E^h_-<0$
(green dotted curve), and hence $n_f(E^h_{-{\bf k}})-n_f(E^e_{-{\bf k}})=1$,
leading to the finite $\rho^s_{\bf q}$ for the outer circle. Whereas,
for $\sigma=1$, there exists the region where the quasiparticle
hole energy is larger than zero ($E^h_+>0$) when $k_z>0$, as shown by blue
dashed curve. This is due to the induced CM momentum, similar to the
conventional FFLO state.\cite{FF} Together with $E^e_+>0$ (brown chain curve), one has
$n_f(E^h_{+{\bf k}})-n_f(E^e_{+{\bf k}})=0$ in this region, 
leading to the open inner circle of the singlet correlation and hence
depairing effect of Cooper pair. 

By using the similar 
analysis, the case with $k_z<0$ can also be understood. The Fermi distributions
of quasiparticle electron and hole in the entire momentum space are plotted in
Figs.~\ref{figyw4}(b) and (c) from full numerical results, respectively. It is
found there exist two arc regions with either quasiparticle 
electron energy below zero or quasiparticle hole one larger than zero, which
exactly correspond to 
the regions with vanishing singlet correlation for the inner circle shown in
Fig.~\ref{figyw3}(c), in consistence with the analysis above. 

Furthermore, it is noted that the emerged FFLO state in our work corresponds to
the gapless one mentioned in the introduction,\cite{FFs-2,FFs-3,FFs-4,FFs-5}
since the gapless quasiparticle energy spectra $|{E^{e/h}_{\bf k}}|=0$
revealed in the gapless FFLO state\cite{FFs-2,FFs-3,FFs-4} 
indicate the emergence of the unpairing regions. By the detailed study
of the SOC dependence (refer to Appendix),  enhanced Pauli limit and hence
enlarged magnetic-field regime of the emerged FFLO state by the SOC is also
observed in our work. We further show that this enhancement of the Pauli
limit is due to the spin-flip terms of the SOC, which suppress the unpairing
regions (also addressed in Appendix).   

\begin{figure}[htb]
  {\includegraphics[width=8.0cm]{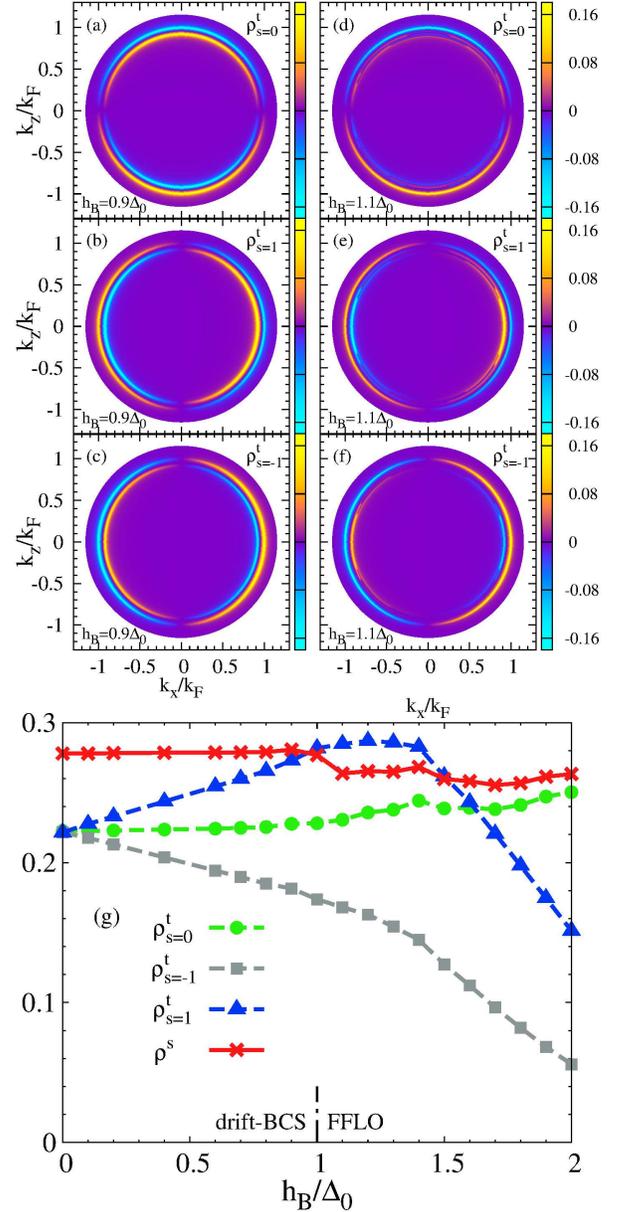}}
  \caption{(a)-(f): momentum dependence of ${\rho}^t_{s=0}$ and
        ${\rho}^t_{s=\pm1}$ in the 
    drift-BCS state at $h_B=0.9\Delta_0$ and FFLO one at $h_B=1.1\Delta_0$
    correspondingly.
    (g): maxima of the singlet and triplet correlations in the
    momentum space as function of $h_B$.  The vertical chain line in (g)
  indicates the crossover between the drift-BCS and FFLO states. }     
\label{figyw5}
\end{figure}

\subsection{Triplet Correlation and Cooper-Pair Spin Polarization}

In this section, by studying the induced $p$-wave triplet correlations in the
pairing regions thanks to the broken space-inversion symmetry by SOC, we show
that the Cooper-pair spin polarization,\cite{TE-0,TE-1,csp1,csp2} which is
predicted to be induced by the 
magnetic field and CM momentum,\cite{csp2} exhibits totally different
magnetic-field dependences in the drift-BCS and FFLO states due to the abrupt changes of the
order parameters and CM momenta. This provides a scheme to
experimentally distinguish these two phases through the reported magnetoelectric Andreev
effect,\cite{csp1,csp2,csp3} in addition to the phase transition.

\subsubsection{Triplet Correlation}
\label{tc}

We first discuss the triplet correlations. Specifically, with the broken
space-inversion symmetry by the SOC, $p$-wave triplet correlations are
induced, plotted in Fig.~\ref{figyw5} at different magnetic fields. From the
figure, it is seen that in the drift-BCS state ($h_B<\Delta_0$), two
separated and complete circles with finite ${\rho}^t_{s=0}$
[Fig.~\ref{figyw5}(a)], ${\rho}^t_{s=1}$ [Fig.~\ref{figyw5}(b)] and
${\rho}^t_{s=-1}$ [Fig.~\ref{figyw5}(c)] are observed in the momentum space,
similar to the singlet case [Fig~\ref{figyw3}(b)]. Moreover, for either the
outer or the inner circle,
it is seen that the triplet correlations show the $p$-wave characters:
${\rho}^t_{s=0}{\propto}h_{{\bf k}z}$ and ${\rho}^t_{s=\pm1}\propto{ih_{{\bf
      k}y}{\mp}h_{{\bf k}x}}$, in agreement with the previous
works.\cite{TE-2,TE-5,TE-10,OP} As for the FFLO state ($h_B>\Delta_0$), as shown in
Figs.~\ref{figyw5}(d),~(e) and~(f), the inner circles of the triplet
correlations are open, similar to the singlet case [Fig~\ref{figyw3}(c)]. Then, the triplet
correlations are only observed in the pairing regions, resembling our
previous work.\cite{OP}    

The magnetic-field dependences of the maximum of the
singlet and triplet correlations in the momentum space are plotted
in Fig.~\ref{figyw5}(g). From the figure, it is seen that in either the
drift-BCS state ($h_B<\Delta_0$) or the FFLO one ($h_B>\Delta_0$), ${\rho}^t_{s=0}$
(dashed curve with dots) and ${\rho}^t_{s=1}$ (dashed curve with triangles)
are comparable to the singlet one (solid curve with crosses) when
$h_B<1.7\Delta_0$. Moreover, it is also noted that
${\rho}^t_{s=1}\ne{\rho}^t_{s=-1}$ when $h_B\ne0$, indicating the generation of
the Cooper-pair spin polarization by the magnetic field
and CM momentum, as predicted in the previous work by Tkachov.\cite{csp2} 

Nevertheless, even with the large $p$-wave triplet correlations (compared with
the singlet one) and the generation of the Cooper-pair spin polarization,
the $p$-wave spin-polarized superfluid is still absent, because of the
vanishing $p$-wave triplet order parameter:
\begin{equation}
\Delta^t({\bf k})=\sum_{\bf k'}V_{\bf k-k'}\rho^t({\bf k'}),
\end{equation}
by the $s$-wave attractive potential $V_{\bf k-k'}=V$ and $p$-wave
character: $\rho^t({\bf k'})=-\rho^t({\bf -k'})$. However, it is proposed
recently in ultracold atom systems\cite{PC-1,PC-2} that in the presence of the
triplet correlation, one can rapidly introduces ${\bf k}$-dependent attractive
potential $V_{\bf k-k'}$ through the Feshbach resonance, and then non-vanishing
$p$-wave superfluid is immediately obtained, at least just after the
introduction of the potential. With this approach, the $p$-wave
spin-polarized superfluid can be expected in the spin-orbit coupled ultracold
atom systems with a magnetic field, according to our results and analysis above.

\begin{figure}[htb]
  {\includegraphics[width=7.5cm]{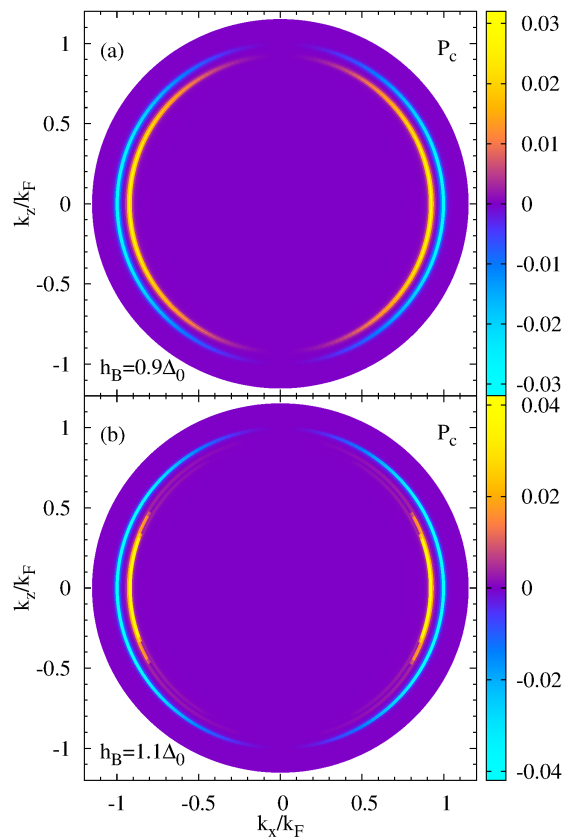}}
\caption{Momentum dependence of Cooper-pair spin polarization in (a)
 the drift-BCS state at $h_B=0.9\Delta_0$ and (b) FFLO one at $h_B=1.1\Delta_0$.}   
\label{figyw6}
\end{figure}

\begin{figure}[htb]
  {\includegraphics[width=8.5cm]{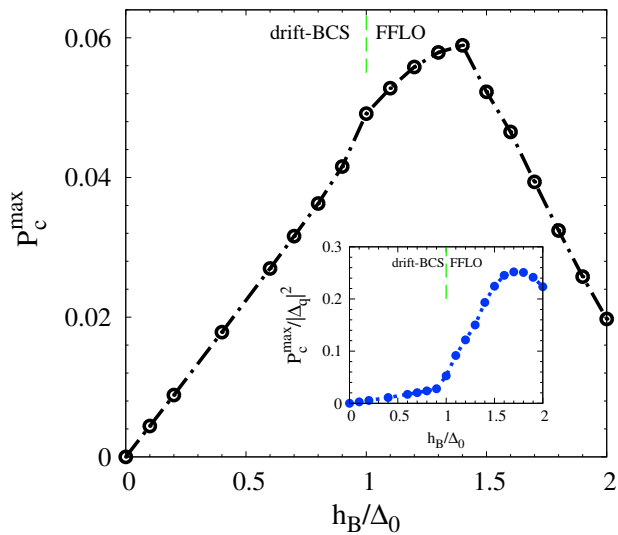}}
\caption{Maximum of Cooper-pair spin polarization in the
  momentum space versus magnetic field. The inset shows $P^{\rm
    max}_c/|\Delta_{\bf q}|^2$ as function of $h_B$. The vertical dashed line
  indicates the crossover between the drift-BCS and FFLO states.}   
\label{figyw7}
\end{figure}

\subsubsection{Cooper-Pair Spin Polarization}
\label{CPSP}

Next, we show that due to the abrupt changes in order parameters and CM momenta
between the drift-BCS and FFLO states, the induced Cooper-pair spin
polarizations mentioned in Sec.~\ref{tc}, exhibit totally different
magnetic-field dependences in
these two phases.

Specifically.  as mentioned in
Sec.~\ref{tc}, with the induced triplet
correlation, the Cooper-pair spin
polarizations, defined as\cite{TE-0,TE-1,csp1,csp2}
\begin{equation}
\label{CSP}
P_c({\bf k})=|{\rho}^t_{s=1}({\bf k})|^2-|{\rho}^t_{s=-1}({\bf k})|^2,
\end{equation}
are induced by the magnetic field, plotted in
Fig.~\ref{figyw6} at different magnetic fields. As seen from the figure, in the
drift-BCS state [Fig.~\ref{figyw6}(a)] and pairing regions of the FFLO
one [Fig.~\ref{figyw6}(b)], it is seen
that $P_c({\bf k})\propto{k^2_x}$, which can be understood from
Eq.~(\ref{tcorrelation}). Specifically, from Eqs.~(\ref{tcorrelation}) and
(\ref{CSP}), one has 
\begin{equation}
P_c({\bf k})\propto|f_{\uparrow\uparrow}(0,{\bf
  k})|^2-|f_{\downarrow\downarrow}(0,{\bf k})|^2=i{\bf f}^t_{\bf q}(0,{\bf
  k})\times{{\bf f}^t_{\bf q}}^*(0,{\bf k})|_z.
\end{equation} 
Then, from Eq.~(\ref{tp}), one immediately finds that 
\begin{equation}
\label{fCSP}
P_c({\bf
  k}){\propto}[{\bf h}_{\bf k}\times({\bf h}_{\bf k}\times{\bm \Omega_{\bf
    q}})]_z|\Delta_{\bf q}|^2=(h^2_{{\bf k}x}+h^2_{{\bf k}y})\Omega_{q{\bf z}}|\Delta_{\bf q}|^2,
\end{equation} in accord with the numerical results (Fig.~\ref{figyw6}). 

The magnetic-field dependence of the maximum of the Cooper-pair spin
polarization $P^{\rm max}_c$ in the momentum space is plotted in
Fig.~\ref{figyw7}. As seen from the figure, with the increase
of $h_B$, in comparison with the linear increase of $P^{\rm max}_c$ in the
drift-BCS state ($h_B<\Delta_0$), {\em nonlinear } increase
of $P^{\rm max}_c$ is observed in the FFLO one when $h_B<1.5\Delta_0$. The
totally different magnetic-field dependences of the Cooper-pair spin 
polarization before and after the phase transition at $h_B=\Delta_0$ provide a scheme to
experimentally distinguish the drift-BCS and FFLO states through the
reported magnetoelectric Andreev effect,\cite{csp1,csp2,csp3} in addition to the
phase transition. Moreover, by further increasing the magnetic field after $1.5\Delta_0$,
it is seen that $P^{\rm max}_c$ markedly decreases and hence a peak, as a
unique feature of the FFLO state, is observed. 

From Eq.~(\ref{fCSP}), the magnetic-field dependence of $P^{\rm max}_c$ can be
clearly understood. Specifically, in the drift-BCS states ($h_B<\Delta_0$), 
with $\Omega_{q{\bf z}}\approx{h_B}$ and the marginal
variation of $\Delta_{\bf q}$ [Fig.~\ref{figyw3}(a)], $P^{\rm max}_c$ increases
linearly with $h_B$. As for the FFLO state, with the increase of $h_B$ at
$\Delta_0<h_B<1.5\Delta_0$, although
$\Delta_{\bf q}$ is suppressed [Fig.~\ref{figyw3}(a)], $\Omega_{q{\bf
    z}}~(\propto=h_B+h_{q{\bf z}})$ is markedly enhanced due to the increased CM
momentum ${\bf q}$ [inset (I) of Fig.~\ref{figyw3}(a)]. 
By the stronger 
enhancement from $\Omega_{q{\bf z}}$ than the suppression from $|\Delta_{\bf
  q}|^2$ in Eq.~(\ref{fCSP}) at $\Delta_0<h_B<1.5\Delta_0$, $P^{\rm max}_c$
increases nonlinearly with $h_B$. This can be justified by plotting
$P^{\rm max}_c/|\Delta_{\bf q}|^2$ ($\propto\Omega_{q{\bf z}}$) versus $h_B$ in
the inset of Fig.~\ref{figyw7}, from which it is seen that $P^{\rm
  max}_c/|\Delta_{\bf q}|^2$ is markedly enhanced at
$\Delta_0<h_B<1.5\Delta_0$. By further increasing $h_B$ after $1.5\Delta_0$,
${\bf q}$ becomes saturated at $0.47q_0{\bf z}$ [inset (I) of 
Fig.~\ref{figyw3}(a)], and hence the suppression of $\Delta_{\bf q}$ leads to
the marked decrease of $P^{\rm max}_c$. 

\section{SUMMARY}

In summary, we have systematically investigated the properties of the
FFLO state with an induced CM momentum in the spin-orbit-coupled
superconductor Li$_2$Pd$_3$B in the presence of the magnetic field. Differing
from the previous theoretical works\cite{FFs-1,FFs-2,FFs-3,FFs-4,FFs-5} where
the study is based on the
numerical calculation of the free-energy minimum with respect to the 
CM momentum and order parameter, in our work, by analytically obtaining the
anomalous Green function and hence the gap equation, the superconducting state
can be determined by computing the energy minimum with respect to
a single parameter, i.e., the CM momentum. Moreover, from the obtained anomalous Green
function, properties of the superconducting state including quasiparticle energy
spectra, singlet and triplet correlations, behaviors of the CM momentum and
order parameter at the phase transition are also 
addressed in our work. 

Specifically, it is found that with the SOC, the CM momentum parallel
to the magnetic field is induced at small magnetic field, similar to the gapped 
FFLO state in the previous works.\cite{FFs-2,FFs-3,FFs-4,FFs-5}
Nevertheless, we have shown that two complete circles of the singlet
correlation due to the SOC are observed in the momentum space, and no unpairing
region with vanishing singlet correlation appears. This is very different from the
conventional FFLO state without SOC, where the CM momentum is induced simultaneously with
the emergence of the unpairing regions.\cite{FF} By further studying the pairing
mechanism, it is shown that the induced CM momentum with SOC at small magnetic
field is due to the energy-spectrum distortion,
resembling the intravalley pairing in graphene\cite{gi} and transition metal
dichalcogenides,\cite{ti} and hence has different origin from the case in conventional
FFLO state.\cite{FF} Therefore, it is more appropriate to refer to such
superconducting state, in which the CM momentum is induced
but no unpairing region is developed, as the drift-BCS state.
By further increasing the magnetic field, abrupt enhancement of the CM momenta
and suppression on the order parameters are observed, indicating the occurrence
of the first-order phase transition. Particularly, we find that open 
circle of the singlet correlation, i.e., unpairing region with vanishing
singlet correlation, is induced after the phase transition, showing the emergence
of the FFLO state. It is further shown that induced unpairing regions arises
from the quasiparticle electron and hole with energies below and larger than
zero, respectively, indicating the emerged FFLO state here corresponds to the
gapless one in the previous works.\cite{FFs-2,FFs-3,FFs-4,FFs-5} Enhanced Pauli
limit and hence enlarged magnetic-field regime of the emerged FFLO state are
also observed in our work, and we demonstrate that the enhancement of the
Pauli limit is due to the spin-flip terms of the SOC, which suppress the
unpairing regions. 

Finally, we discuss the triplet correlation induced by the
SOC, and show that in the presence of the triplet correlation, 
the Cooper-pair spin polarizations,\cite{TE-0,TE-1,csp1,csp2} induced by the
magnetic field and CM momentum, exhibit totally different magnetic-field
dependences between the drift-BCS and FFLO states. This difference between the
drift-BCS and FFLO states, arising from the abrupt 
changes in order parameters and CM momenta, provides a scheme to
experimentally distinguish these two phases
through the reported magnetoelectric Andreev effect,\cite{csp1,csp2,csp3} in
addition to the phase transition.

\begin{figure}[htb]
  {\includegraphics[width=8.5cm]{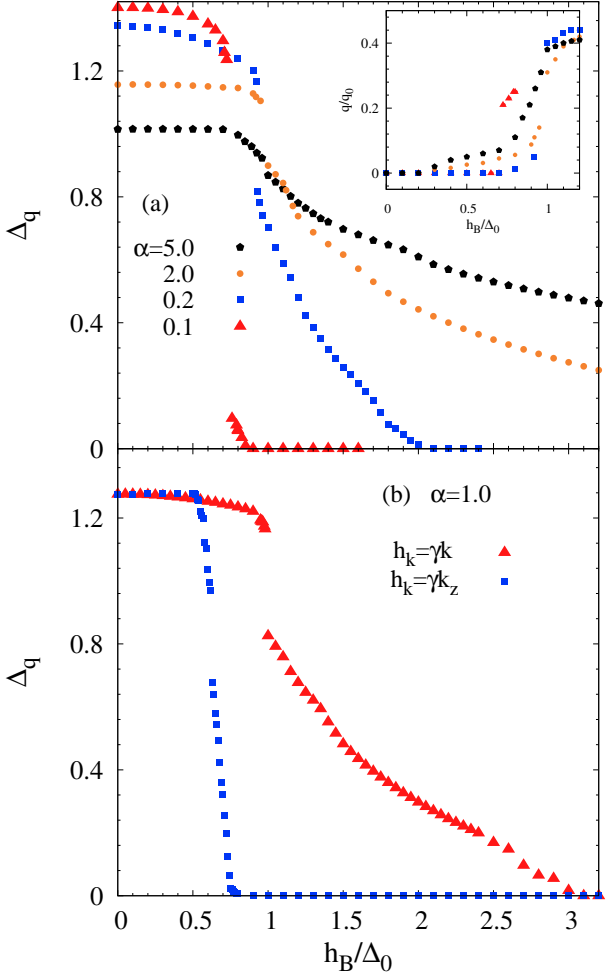}}
\caption{(a): $\Delta_{\bf q}$ versus $h_B$ at different SOCs. The inset
  shows ${\bf q}=q{\bf z}$ as function of $h_B$ at different
  SOCs. In the calculation, $\gamma=\alpha\gamma_0$ with $\gamma_0$ being the SOC strength in
  Li$_2$Pd$_3$B. (b): $\Delta_{\bf q}$ versus $h_B$ at $\alpha=1$. Triangles
  (Squares) in (b): the spin-flip terms of the SOC are included (removed) by
  setting ${\bf h}_{\bf k}=\gamma{\bf k}$ (${\bf h}_{\bf k}=\gamma{k_z}{\bf z}$).}    
\label{figyw8}
\end{figure}

\begin{appendix}

\section{SOC DEPENDENCE}

In this part, we address the SOC dependence of the superconducting state. The
magnetic field dependences of the order parameter $\Delta_{\bf 
  q}$ and CM momentum ${\bf q}=q{\bf z}$ are plotted in
Fig.~\ref{figyw8}(a) and the inset of the same figure at different SOCs,
respectively. As seen from the figure,  with the increase of the SOC strength, 
the Pauli limit is enhanced and hence the regime where the FFLO occurs is
enlarged, in accord with the previous experiments\cite{ev1,ev2,ev3,ev4} and
prediction.\cite{FFs-1}

The enhancement of the Pauli limit is due to the
spin-flip terms of the SOC (perpendicular to ${\bf h}_B$).
This can be seen from Fig.~\ref{figyw8}(b) where we plot $\Delta_{\bf q}$
versus $h_B$ with and without the spin-flip terms of the SOC. From the figure,
it is seen that in comparison with the case at ${\bf h}_{\bf k}=\gamma{\bf k}$
(triangles), in the situation without the spin-flip terms of the SOC (${\bf
  h}_{\bf k}=\gamma{k_z}{\bf z}$), as shown by squares, the Pauli limit is
markedly suppressed and the FFLO state occurs in a narrow magnetic-field regime.
This is because that away from the $k_z$ axis,
the spin-flip terms of the SOC couple the quasiparticle
electrons $E^e_{+{\bf k}}$ and $E^e_{-{\bf k}}$ (holes $E^h_{+{\bf k}}$ and
$E^h_{-{\bf k}}$) with different spin polarizations in Fig.~\ref{figyw4}(a).
Consequently, due to $E^e_{+{\bf k}}>0$ ($E^h_{-{\bf 
    k}}<0$) shown in Fig.~\ref{figyw4}(a), the regions with $E^e_{-{\bf k}}<0$
($E^h_{+{\bf k}}>0$) mentioned in Sec.~\ref{pd}, i.e., the unpairing regions, 
are suppressed, leading to the suppressed Pauli limit.

Furthermore, it is noted that at very small SOC 
$\gamma=0.1\gamma_0$ (shown by triangles), the FFLO state occurs in a narrow 
regime $0.7\Delta_0<h_B<0.8\Delta_0$, close to the conventional FFLO one 
without SOC ($0.66\Delta_0<h_B<0.8\Delta_0$).\cite{FF} With the increase of the
SOC, the variation of $\Delta_{\bf q}$ at the phase transition between the
drift-BCS and FFLO states is suppressed. Particularly, when we extend to a large
SOC $\gamma=5\gamma_0$ [shown by pentagons in Fig.~\ref{figyw8}(a)], the
variation of $\Delta_{\bf q}$ at the phase transition ($h_B=\Delta_0$) becomes
nearly indistinguishable but still exists. This case is very similar to Fig.~4(b) in
Ref.~\onlinecite{FFs-2}. This suppressed variation of $\Delta_{\bf
  q}$ at the phase transition is due to the
enhancement of the CM momentum in the drift-BCS state by the SOC, leading to the
close CM momenta between the drift-BCS ($h_B<\Delta_0$) and FFLO
($h_B>\Delta_0$) states, as shown by pentagons in the inset of
Fig.~\ref{figyw8}(a).    

Nevertheless, with the larger SOC $\gamma>5\gamma_0$
($\Delta_0\ll\gamma{k_F}$), the mean-field theory $\Delta_{\bf
  q}=-V\sum'_{\bf k}\langle{\phi_{\uparrow{\bf k+q}}\phi_{\downarrow{\bf
      -k+q}}}\rangle$ in colinear space is inappropriate due to
the large spin-flip terms of the SOC. In this case, pairing mechanism and
the generation of the CM momentum should be discussed in the helix space, which is
beyond the scope of our work. 

\end{appendix}


\begin{thebibliography}{0}


\bibitem{BCS} J. Bardeen, L. N. Cooper, and J. R. Schrieffer, Phys. Rev. {\bf
    106}, 162 (1957).
\bibitem{TE-0} A. J. Leggett, Rev. Mod. Phys. {\bf 47}, 331 (1975).
\bibitem{TE-1} M. Sigrist and K. Ueda, Rev. Mod. Phys. {\bf 63}, 239 (1991).
\bibitem{TE-2} L. P. Gor'kov and E. I. Rashba, Phys. Rev. Lett. {\bf 87},
  037004 (2001). 
\bibitem{TE-3} P. A. Frigeri, D. F. Agterberg, A. Koga, and M. Sigrist, Phys.
Rev. Lett. {\bf 92}, 097001 (2004).
\bibitem{TE-4} Z. H. Yang, J. Wang, and K. S. Chan,
  Supercond. Sci. Technol. {\bf 22}, 055012 (2009).   
\bibitem{TE-5} E. Bauer and M. Sigrist, {\em Non-centrosymmetric
  Superconductors: Introduction and Overview} (Springer Science and Business
  Media, Berlin, 2012). 
\bibitem{TE-6} X. Liu, J. K. Jain, and C. X. Liu, Phys. Rev. Lett. {\bf 113},
  227002 (2014).
\bibitem{TE-7} C. R. Reeg and D. L. Maslov, Phys. Rev. B {\bf 92}, 134512 (2015).
\bibitem{TE-8} C. Triola, D. M. Badiane, A. V. Balatsky, and E. Rossi,
  Phys. Rev. Lett. {\bf 116}, 257001 (2016).
\bibitem{TE-10} T. Yu and M. W. Wu, Phys. Rev. B {\bf 93}, 195308 (2016). 
\bibitem{OP} F. Yang and M. W. Wu,  Phys. Rev. B {\bf 95}, 075304 (2017). 

\bibitem{Sr-1} P. W. Anderson and W. F. Brinkman, Phys. Rev. Lett. {\bf 30},
  1108 (1973).
\bibitem{Sr-2} W. F. Brinkman, J. W. Serene, and P. W. Anderson, Phys. Rev. A
  {\bf 10}, 2386 (1974).
\bibitem{Sr-3} T. M. Rice and M. Sigrist, J. Phys. Condens. Matter {\bf 7}, L643
(1995).
\bibitem{Sr-4} Y. Maeno, H. Hashimoto, K. Yoshida, S. Nishizaki, T. Fujita,
J. G. Bednorz, and F. Lichtenberg, Nature (London) {\bf 372}, 532 (1994).
\bibitem{Sr-5} K. Ishida, Y. Kitaoka, K. Asayama, S. Ikeda, S. Nishizaki, Y.
Maeno, K. Yoshida, and T. Fujita, Phys. Rev. B {\bf 56}, R505(R) (1997).
\bibitem{Sr-6} N. Read and D. Green, Phys. Rev. B {\bf 61}, 10267 (2000).
\bibitem{Sr-7} D. A. Ivanov, Phys. Rev. Lett. {\bf 86}, 268 (2001).
\bibitem{Sr-8} A. P. Mackenzie and Y. Maeno, Rev. Mod. Phys. {\bf 75}, 657
  (2003). 

\bibitem{SP-1} R. Meservey and P. M. Tedrow, Phys. Rep. {\bf 238}, 173 (1994).
\bibitem{SP-2} {\em Semiconductor Spintronics and Quantum Computation}, edited
by D. D. Awschalom, D. Loss, and N. Samarth (Springer, Berlin, 2002).
\bibitem{SP-3} I. \u{Z}uti{\'c}, J. Fabian, and S. D. Sarma, Rev. Mod. Phys. {\bf
    76}, 323 (2004).
\bibitem{SP-4} A. I. Buzdin, Rev. Mod. Phys. 77, 935 (2005).
\bibitem{SP-5} F. S. Bergeret, A. F. Volkov, and K. B. Efetov, Rev. Mod. Phys.
{\bf 77}, 1321 (2005).
\bibitem{SP-6} J. Fabian, A. M. Abiague, C. Ertler, P. Stano, and I. \u{Z}uti{\'c},
  Acta Phys. Slov. {\bf 57}, 565 (2007). 
\bibitem{SP-7} M. W. Wu, J. H. Jiang, and M. Q. Weng, Phys. Rep. {\bf 493}, 61 (2010).
\bibitem{SP-8} {\em Handbook of Spin Transport and Magnetism}, edited by
  E. Y. Tsymbal and I. \u{Z}uti{\'c} (CRC, Boca Raton, FL, 2011). 
\bibitem{SP-9} M. Eschrig, Phys. Today {\bf 64}(1), 43 (2011).
\bibitem{SP-10} J. Linder and J. W. A. Robinson, Nat. Phys. {\bf 11}, 307 (2015).


\bibitem{FF} P. Fulde and A. Ferrell, Phys. Rev. {\bf 135}, A550 (1964).
\bibitem{LO} A. I. Larkin and Y. N. Ovchinnikov, Sov. Phys. JETP {\bf 20}, 762
  (1965) [Zh. Eksp. Teor. Fiz. {\bf 47}, 1136 (1964)].


\bibitem{FF-od1} K. Maki, Phys. Rev. {\bf 148}, 362 (1966).
\bibitem{FF-od2} L. Gruenberg and L. Gunther, Phys. Rev. Lett. {\bf 16}, 996 (1966).

\bibitem{hf-1} H. A. Radovan, N. A. Fortune, T. P. Murphy, S. T. Hannahs, E. C.
Palm, S. W. Tozer, and D. Hall, Nature {\bf 425}, 51 (2003). 
\bibitem{hf-2} A. Bianchi, R. Movshovich, C. Capan, P. G. Pagliuso, and
J. L. Sarrao, Phys. Rev. Lett. {\bf 91}, 187004 (2003). 
\bibitem{hf-3} C. Capan, A. Bianchi, R. Movshovich, A. D. Christianson,
A. Malinowski, M. F. Hundley, A. Lacerda, P. G. Pagliuso, and J. L. Sarrao,
Phys. Rev. B {\bf 70}, 134513 (2004).  
\bibitem{hf-4} T. Watanabe, Y. Kasahara, K. Izawa, T. Sakakibara, Y. Matsuda, C.
J. van der Beek, T. Hanaguri, H. Shishido, R. Settai, and Y. Onuki,
Phys. Rev. B {\bf 70}, 020506 (2004).
\bibitem{hf-5} K. Kakuyanagi, M. Saitoh, K. Kumagai, S. Takashima, M. Nohara,
  H. Takagi, and Y. Matsuda, Phys. Rev. Lett. {\bf 94}, 047602 (2005). 
\bibitem{hf-6} K. Kumagai, M. Saitoh, T. Oyaizu, Y. Furukawa, S. Takashima, M.
Nohara, H. Takagi, and Y. Matsuda, Phys. Rev. Lett. {\bf 97}, 227002 (2006).
\bibitem{hf-7} V. F. Mitrovic, M. Horvatic, C. Berthier, G. Knebel, G. Lapertot,
  and J. Flouquet, Phys. Rev. Lett. {\bf 97}, 117002 (2006).
\bibitem{hf-8} Y. Matsuda and H. Shimahara, J. Phys. Soc. Jpn. {\bf 76}, 051005
  (2007). 
\bibitem{hf-9} M. Kenzelmann, S. Gerber, N. Egetenmeyer, J. L. Gavilano,
  Th. Str{\" a}ssle, A. D. Bianchi, E. Ressouche, R. Movshovich, E. D. Bauer,
  J. L. Sarrao, and J. D. Thompson, Phys. Rev. Lett. {\bf 104}, 127001 (2010).

\bibitem{ca-1} R. Casalbuoni and G. Nardulli, Rev. Mod. Phys. {\bf 76}, 263
  (2004).
\bibitem{ca-2} T. Mizushima, K. Machida, and M. Ichioka, Phys. Rev. Lett. {\bf
    94}, 060404 (2005). 
\bibitem{ca-3} M. W. Zwierlein, A. Schirotzek, C. H. Schunck, and W. Ketterle,
Science {\bf 311}, 492 (2006).
\bibitem{ca-4} D. E. Sheehy and L. Radzihovsky, Phys. Rev. Lett. {\bf 96},
  060401 (2006).  
\bibitem{ca-5} M. M. Parish, S. K. Baur, E. J. Mueller, and D. A. Huse,
  Phys. Rev. Lett. {\bf 99}, 250403 (2007). 
\bibitem{ca-6} T. K. Koponen, T. Paananen, J. P. Martikainen, and P. T{\"
    o}rm{\" a}, Phys. Rev. Lett. {\bf 99}, 120403 (2007).
\bibitem{ca-7} Y. A. Liao, A. S. C. Rittner, T. Paprotta, W. Li,
  G. B. Partridge, R. G. Hulet, S. K. Baur, and E. J. Mueller, Nature (London)
  {\bf 467}, 567 (2010).
\bibitem{ca-8} F. Chevy and C. Mora, Rep. Prog. Phys. {\bf 73}, 112401 (2010).  
\bibitem{ca-9} Z. Cai, Y. Wang, and C. Wu, Phys. Rev. A {\bf 83}, 063621
  (2011). 


\bibitem{Fe1} A. Gurevich, Phys. Rev. B {\bf 82}, 184504 (2010).
\bibitem{Fe2} K. Cho, H. Kim, M. A. Tanatar, Y. J. Song, Y. S. Kwon,
  W. A. Coniglio, C. C. Agosta, A. Gurevich, and R. Prozorov, Phys. Rev. B {\bf
    83}, 060502(R) (2011). 
\bibitem{Fe3} S. Khim, B. Lee, J. W. Kim, E. S. Choi, G. R. Stewart, and K. H. Kim,
  Phys. Rev. B {\bf 84}, 104502 (2011).
\bibitem{Fe4} A. Gurevich, Rep. Prog. Phys. {\bf 74}, 124501 (2011).



\bibitem{os-0} H. Shimahara, J. Phys. Soc. Jpn. {\bf 66}, 541 (1997).
\bibitem{os-1} M. A. Tanatar, T. Ishiguro, H. Tanaka, and H. Kobayashi,
  Phys. Rev. B {\bf 66}, 134503 (2002).
\bibitem{os-2} S. Uji, T. Terashima, M. Nishimura, Y. Takahide, T. Konoike,
  K. Enomoto, H. Cui, H. Kobayashi, A. Kobayashi, H. Tanaka, M. Tokumoto,
  E. S. Choi, T. Tokumoto, D. Graf, and J. S. Brooks, Phys. Rev. Lett. {\bf 97},
  157001 (2006). 
\bibitem{os-3} R. Lortz, Y. Wang, A. Demuer, P. H. M. B{\" o}ttger, B. Bergk,
  G. Zwicknagl, Y. Nakazawa, and J. Wosnitza, Phys. Rev. Lett. {\bf 99}, 187002
  (2007).
\bibitem{os-4} B. Bergk, A. Demuer, I. Sheikin, Y. Wang, J. Wosnitza,
  Y. Nakazawa, and R. Lortz, Phys. Rev. B {\bf 83}, 064506 (2011). %
\bibitem{os-5} J. A. Wright, E. Green, P. Kuhns, A. Reyes, J. Brooks,
  J. Schlueter, R. Kato, H. Yamamoto, M. Kobayashi, and S. E. Brown,
  Phys. Rev. Lett. {\bf 107}, 087002 (2011).
\bibitem{os-6} R. Beyer and J. Wosnitza, Low Temp. Phys. {\bf 39}, 225 (2013).
\bibitem{os-7} H. Mayaffre, S. Kr{\"a}mer, M. Horvatic, C. Berthier,
  K. Miyagawa, K. Kanoda, and V. V.  Mitrovic, Nat. Phys. {\bf 10}, 928 (2014).
  
\bibitem{scatt-1} L. G. Aslamazov, Sov. Phys. JETP {\bf 28}, 773 (1969)
  [Zh. Eksp. Teor. Fiz. {\bf 55}, 1477 (1968)].
\bibitem{scatt-2}  S. Takada, Prog. Theor. Phys. {\bf 43}, 27 (1970).


\bibitem{Od-1} L. W. Gruenberg and L. Gunther, Phys. Rev. Lett. {\bf 16}, 996 (1966).
\bibitem{Od-2} H. Adachi and R. Ikeda, Phys. Rev. B {\bf 68}, 184510 (2003).


\bibitem{ap-1} H. Shimahara, Phys. Rev. B {\bf 50}, 12760 (1994).

\bibitem{SOC-1} G. Dresselhaus, Phys. Rev. {\bf 100}, 580 (1955).
\bibitem{SOC-2} Y. A. Bychkov and E. I. Rashba, J. Phys. C {\bf 17}, 6039 (1984).
\bibitem{SOC-3} Y. A. Bychkov, JETP Lett. {\bf 39}, 78 (1984).

\bibitem{ap-2} L. Dong, L. Jiang, H. Hu, and J. Pu, Phys. Rev. A {\bf 87},
  043616 (2013).



  
\bibitem{FFs-1} Z. Zheng, M. Gong, X. Zou, C. Zhang, and G. C. Guo, Phys. Rev. A
  {\bf 87}, 031602 (2013). 
\bibitem{FFs-2} L. Dong, L. Jiang, and H. Pu, New J. Phys. {\bf 15}, 075014
  (2013). 
\bibitem{FFs-3} X. J. Liu and H. Hu, Phys. Rev. A {\bf 87}, 051608(R)
  (2013). 
\bibitem{FFs-4} F. Wu, G. C. Guo, W. Zhang, and W. Yi, Phys. Rev. Lett. {\bf
    110}, 110401 (2013).
\bibitem{FFs-5} X. F. Zhou, G. C. Guo, W. Zhang, and W. Yi, Phys. Rev. A {\bf 87}, 063606 (2013).
\bibitem{FFs-6} C. Chen, Phys. Rev. Lett. {\bf 111}, 235302 (2013).
\bibitem{FFs-7} Y. Xu, C. L. Qu, M. Gong, and C. W. Zhang, Phys. Rev. A {\bf 89}, 013607
  (2014). 
\bibitem{FFs-8} G. Zwicknagl, S. Jahns, and P. Fulde, arXiv:1701.09121.

\bibitem{ev1} D. Aoki, A. Huxley, E. Ressouche, D. Braithwaite, J. Flouquet,
  J. P. Brison, E. Lhotel, and C. Paulsen, Nature {\bf 413}, 613 (2001).
\bibitem{ev2} M. B. Shalom, M. Sachs, D. Rakhmilevitch, A. Palevski, and
  Y. Dagan, Phys. Rev. Lett. {\bf 104}, 126802 (2010). 
\bibitem{ev3} J. F. Mercure, A. F. Bangura, X. F. Xu, N. Wakeham, A. Carrington,
  P. Walmsley, M. Greenblatt, and N. E. Hussey,  Phys. Rev. Lett. {\bf 108},
  187003 (2012).
\bibitem{ev4} M. Smidman, M. B. Salamon, H. Q. Yuan, and D. F. Agterberg,
  Rep. Prog. Phys. {\bf 80}, 036501 (2017). 


\bibitem{Li1} K. W. Lee and W. E. Pickett, Phys. Rev. B {\bf 72}, 174505
  (2005). 
\bibitem{Li2} H. Takeya, K. Hirata, K. Yamaura, and K. Togano, M. E. Massalami,
  R. Rapp, and F. A. Chaves, B. Ouladdiaf, Phys. Rev. B {\bf 72}, 104506
  (2005). 
\bibitem{Li3} R. Khasanov, I. L. Landau, C. Baines, F. L. Mattina,
  A. Maisuradze, K. Togano, and H. Keller, Phys. Rev. B {\bf 73}, 214528
  (2006). 
\bibitem{Li4} S. Tsuda, T. Yokoya, T. Kiss, T. Shimojima, K. Ishizaka, S. Shin,
  T. Togashi, S. Watanabe, C. Q. Zhang, C. T. Chen, I. Hase, H. Takeya,
  K. Hirata, and K. Togano, J. Phys. Soc. Jpn. {\bf 78}, 034711 (2009). 
\bibitem{Li5} S. P. Mukherjee and T. Takimoto, Phys. Rev. B {\bf 86}, 134526
  (2012). 
\bibitem{Li6} H. Q. Yuan, D. F. Agterberg, N. Hayashi, P. Badica,
  D. Vandervelde, K. Togano, M. Sigrist, and M. B. Salamon,
  Phys. Rev. Lett. {\bf 97}, 017006 (2016). 

\bibitem{gi} M. Einenkel and K. B. Efetov, Phys. Rev. B {\bf 84}, 214508 (2011).
\bibitem{ti} S. Tsuchiya, J. Goryo, E. Arahata, and M. Sigrist, Phys. Rev. B
  {\bf 94}, 104508 (2016). 

\bibitem{csp1} A. D. Hillier, J. Quintanilla, B. Mazidian, J. F. Annett, and
  R. Cywinski, Phys. Rev. Lett. {\bf 109}, 097001 (2012).
\bibitem{csp2} G. Tkachov, Phys. Rev. Lett. {\bf 118}, 016802 (2017).
\bibitem{csp3} P. H{\" o}gl, A. M. Abiague, I. {\v Z}uti{\' c}, and J. Fabian,
  Phys. Rev. Lett. {\bf 115}, 116601 (2015).

\bibitem{G1} A. A. Abrikosov, L. P. Gorkov, and I. E. Dzyaloshinski,
{\em Methods of Quantum Field Theory in Statistical Physics} (Prentice Hall,
Englewood Cliffs, NJ, 1963). 
\bibitem{G2} A. L. Fetter and J. D. Walecka, {\em Quantum Theory of Many
    Particle Systems} (McGraw-Hill, New York, 1971).
\bibitem{G3} G. D. Mahan, {\em Many Particle Physics} (Plenum, New York, 1990).

\bibitem{Gorkov} L. P. Gor’kov, Zh. Eksp. Teor. Fiz. {\bf 36}, 1918 (1959)
  [Sov. Phys. JETP {\bf 9}, 1364 (1959)]; Zh. Eksp. Teor. Fiz. {\bf 37}, 1407
  (1959) [Sov. Phys. JETP {\bf 10}, 998 (1960)].

\bibitem{tp1} G. Annunziata, D. Manske, and J. Linder, Phys. Rev. B {\bf 86},
174514 (2012).
\bibitem{tp2} D. Fritsch and J. F. Annett, J. Phys. Condens. Matter {\bf 26},
274212 (2014).
\bibitem{tp3} T. Yu and M. W. Wu, Phys. Rev. B {\bf 94}, 205305 (2016).

\bibitem{PC-1} Y. Endo, D. Inotani, R. Hanai, and Y. Ohashi, Phys. Rev. A {\bf
    92}, 023610 (2015). 
\bibitem{PC-2} T. Yamaguchi and Y. Ohashi, Phys. Rev. A. {\bf 92}, 013615
  (2015).
  
\end{thebibliography}
\end{document}